\documentclass[utf8]{beilstein}

\usepackage{amsmath,amssymb}
\usepackage{pifont}
\usepackage{hyperref}

\providecommand{\U}[1]{\protect\rule{.1in}{.1in}}
\renewcommand{\epsilon}{{\varepsilon}}

\nolinenumbers

\begin{document}
\title{Experimental and theoretical study of field dependent spin splitting at ferromagnetic-insulator - superconductor interfaces}
\author{P. Machon}
\affiliation{Department of Physics, University of Konstanz, D-78457 Konstanz, Germany}
\author{M. J. Wolf}
\affiliation{Institute of Nanotechnology, Karlsruhe Institute of Technology (KIT), D-76021 Karlsruhe, Germany}
\affiliation{present address: Institute for Technical Physics, Karlsruhe Institute of Technology (KIT), D-76021 Karlsruhe, Germany}
\author*{D. Beckmann}{Detlef.Beckmann@kit.edu}
\affiliation{Institute for Quantum Materials and Technologies, Karlsruhe Institute of Technology (KIT), D-76021 Karlsruhe, Germany}
\author*[1]{W. Belzig}{Wolfgang.Belzig@uni-konstanz.de}
\maketitle

\begin{abstract}
We present a combined experimental and theoretical work that investigates the magnetic proximity effect at a ferromagnetic-insulator - superconductor (FI-S) interface. The simulation is based on the boundary condition for diffusive quasiclassical Greens functions, that accounts for arbitrarily strong spin-dependent effects and spin-mixing angles. The experimentally found differential conductance of an EuS-Al heterostructure is compared with a theoretical calculation. With the assumption of a uniform spin-mixing angle that depends on the externally applied field, we find good agreement between theory and experiment. The theory depends only on very few parameters, mostly specified by the experimental setup. We determine the effective spin of the interface moments as $ J\approx 0.74\hbar $.
\end{abstract}

\section{Introduction}
The proximity effect between superconductors and ferromagnets has been investigated intensively in recent years \cite{buzdin:05,bergeret:05}, giving rise to the field of superconducting spintronics \cite{linder2015,eschrig2015}.
Ferromagnetic insulators like EuO and EuS are interesting materials since they show ferromagnetism (almost ideal Heisenberg ferromagnet), but are electrically insulating at the same time \cite{matthias:61,mcguire:62,mairoser:15}. Magnetic insulators have been used successfully, e.g., in magnetic Josephson junctions \cite{pal2014}, superconducting spin switches \cite{li2013b}, and for studying the triplet proximity effect \cite{diesch2018}.
Ferromagnetic insulators are a good probe of the spin-dependent proximity effect in bilayer structures due to the reduced number of free parameters. In turn, this kind of junctions also provide information on the details of the internal magnetization behavior of ferromagnetic insulators in an external field. To be specific, in a simple stacked structure one observes the proximity effect that solely depends on the internal spin-degrees of freedom (spin mixing angles \cite{tokuyasu:88}), since the conductance is zero, in contrast to a metallic ferromagnet. The absence of conductance related parameters (transmission and polarization of each channel) strongly simplifies the  boundary condition to a ferromagnetic insulator \cite{machon:15,Eschrig_Cottet_Belzig_Linder_2015}, which has in the meanwhile been extended to insulating antiferromagnets \cite{Kamra_Rezaei_Belzig_2018}. Thus, one has the opportunity to quantitatively study the microscopic mechanisms that influence the superconducting density of states, in a way that it mainly shifts and spin-splits the peaks at the superconducting gap edge. Such shifts and following possibility to create of Shiba bands \cite{Shiba_1968,Zittartz_Bringer_Mueller-Hartmann_1972} have been investigated theoretically also recently  in related systems \cite{Ouassou_Pal_Blamire_Eschrig_Linder_2016,Belzig_Beckmann_2017}.

\begin{figure}
\centering
(a)\includegraphics[width=0.45\linewidth]{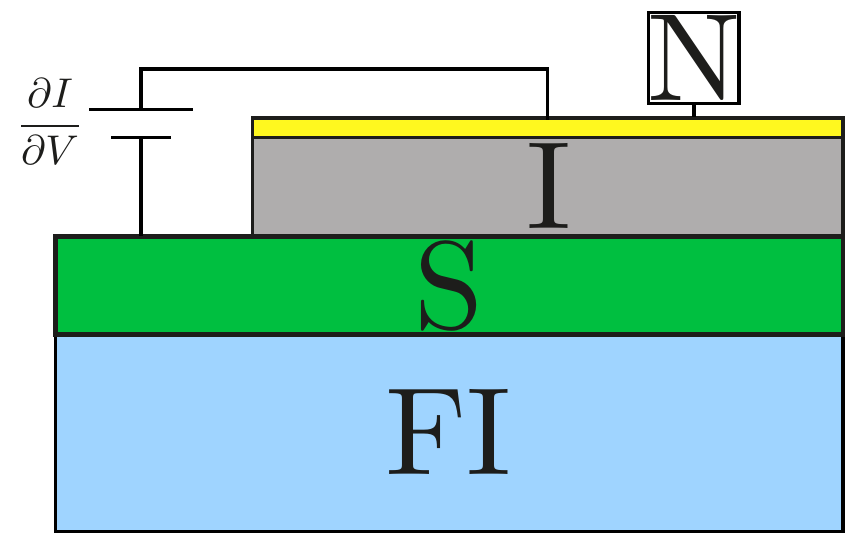}
(b)\includegraphics[width=0.3\linewidth]{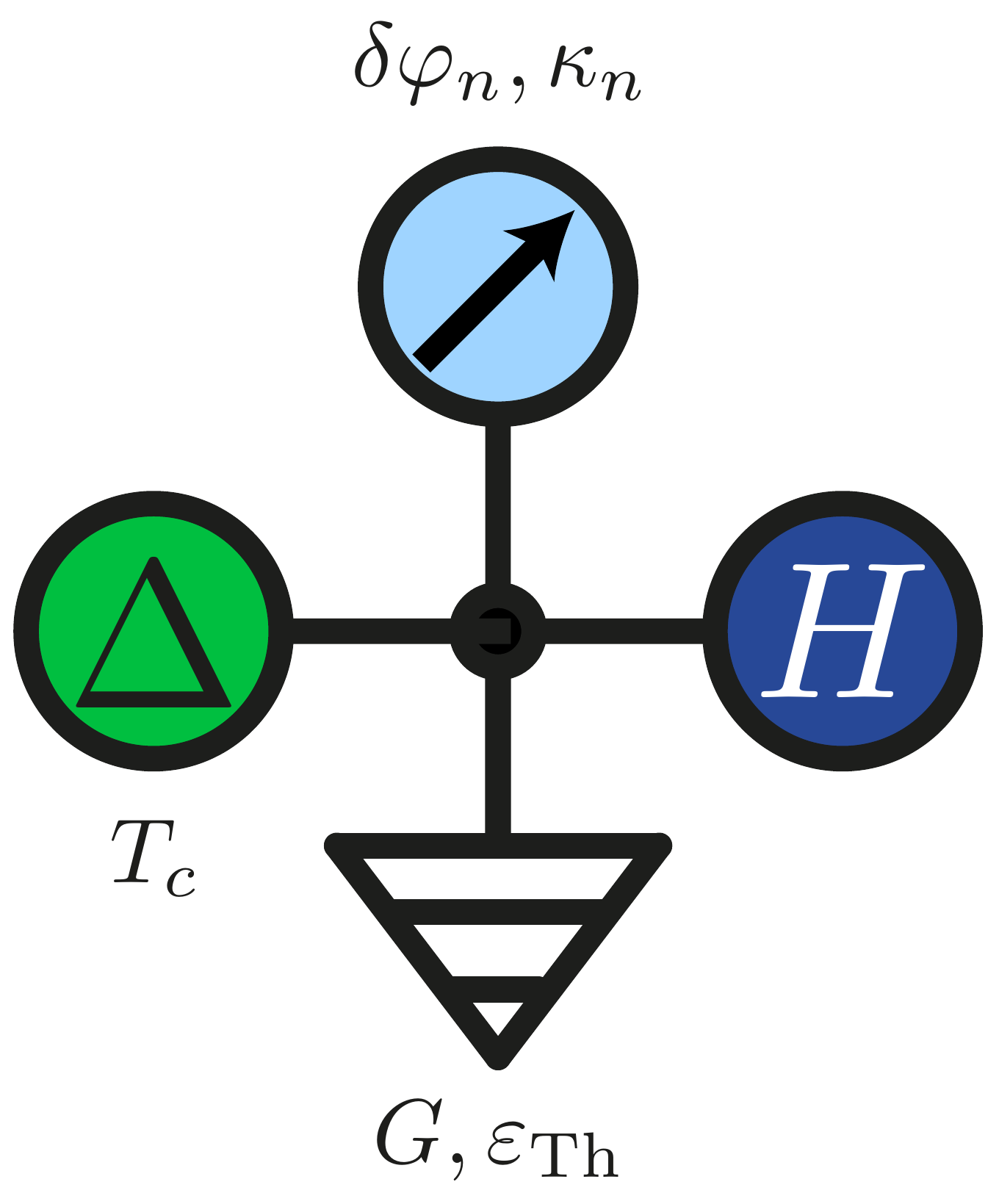}
\caption{\label{sys_fis}
(a) The experimental setup of the FI-S bilayer. The differential conductance is measured with help of the top normal contact (N) that is separated by a thick insulating barrier (I). (b) Circuit diagram to represent the FI-S bilayer in the quantum circuit theory. The superconductor is represented by the node, the $\Delta$-source term, the $\varepsilon_{\rm Th}$-leakage term and it's normal state conductance $G$. The ferromagnetic insulator adds the $\delta\varphi_n$-term with magnetization directions $\kappa_n$ for each channel and in accordance with the experiment an external exchange field is added, here represented by the $H$-pseudo-terminal.}
\end{figure}

\begin{figure}
\centering
\includegraphics[width=0.9\linewidth]{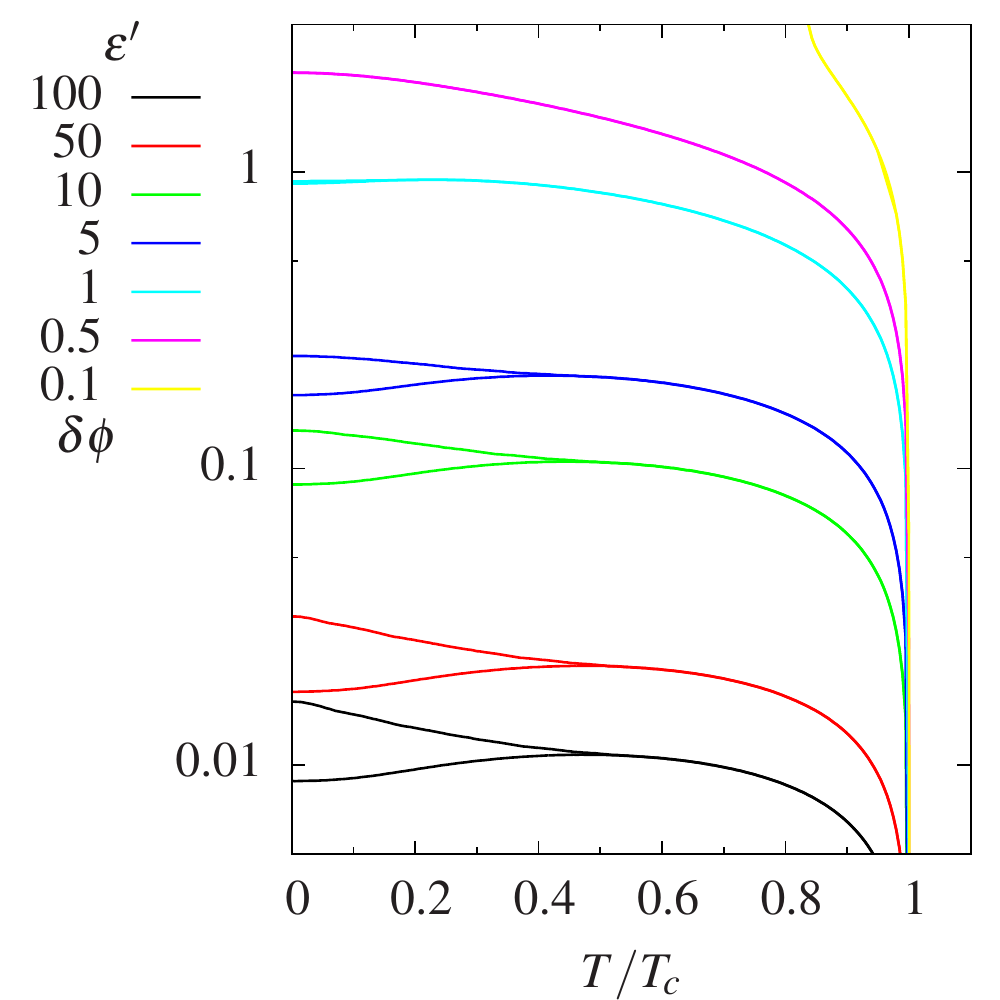}
\caption{\label{gap_fis}
Phase diagram of a superconductor in proximity to a ferromagnetic insulator in dependence on the temperature and the spin mixing angle at the interface to the FI, for values of $\varepsilon'=[100,0.1]$.}
\end{figure}

In principle this is not a new story and has been discussed e.g. in \cite{tokuyasu:88} in the clean and in \cite{cottet:09} in the dirty limit. In our particular case, we want to describe experimental findings in an EuS-Al heterostructure, we thus work in the dirty limit that is known to sufficiently describe Al-superconductors. What is especially new here, is that we can treat the spin mixing not only to linear (Zeeman-type splitting) or second order (pair breaking due to spin-dependent scattering) in the related phase shifts ($\delta\phi$), but exactly. Thus the distribution of spin mixing angles ($\delta\phi_n$, $n$ is the channel index) along the transport channels is the only unknown in the theory. This distribution strongly depends on the applied external field, but so far no theory exists that relates these two quantities. Here, we show how an assumed distribution can directly be probed in a fully electronic experiment, measuring the differential conductance of the aluminum stripe. For simplicity we will only use a singular distribution with a single spin-mixing angle for which the density of states can in principal be calculated analytically, but the theory is not restricted to this case. However, even with this simple model we can reproduce the experimental data to a high extent, that is not so well described in other theories like e.g. \cite{tedrow:71,cottet:09}. E.g Ref.~\cite{tedrow:71} takes into account the Zeeman-splitting of the superconducting density of states induced by the external field, and \cite{cottet:09} incorporate spin mixing angles but only up to the forth order.

The setup of the underlying experiment is shown in fig.~\ref{sys_fis}~(a), it consists (bottom-up) of an EuS substrate, a superconducting (Al) stripe, and a normal metal stripe that is separated from the superconductor by an oxide film acting as the probe electrode to measure the differential conductance of the superconductor and is therefore assumed not to influence the system properties. Since the size of the ``detector'' electrode is not small (unlike the tip of a scanning tunneling microscope) and the FI proximizes the whole superconductor, we assume that the magnetization can be modeled by one magnetization direction, that results from averaging over the internal magnetic structure. In the language of the circuit theory \cite{nazarov:99} this means that we can reproduced the whole system with a single node as depicted in fig.~\ref{sys_fis}~(b). Precisely, the superconductor is represented by the node that has a ``source'' (of coherence) term (marked with $\Delta$), another pseudo terminal that models the spin mixing angles ($\delta\phi$) induced by the FI, the external field ($H$), and the ``leakage'' (of coherence) term that is characterized by the Thouless energy of the superconductor ($\varepsilon_{\rm Th}$) and it's normal state conductance $G$.

\section{Model}

To describe the FI-S bilayer as illustrated in fig.~\ref{sys_fis}~(b) within the circuit theory \cite{nazarov:99}, we use the formalism for the boundary conditions for spin-dependent connectors developed in \cite{machon:15} which agrees with the results of \cite{Eschrig_Cottet_Belzig_Linder_2015}. This boundary condition for the Usadel equation was derived and shown how this BC can be applied to a ferromagnetic insulator - superconductor bilayer system. Defining the spin-dependent Green's function of the superconductor as $\hat G_{\sigma}=\sum_{i=1}^3g_{i,\sigma}\hat\tau_i$ one finds $g_2=0$. Here, $ \sigma=\pm $ denotes the spin index. We obtain the following equation that determines the Green's function of the superconductor:
\begin{align}
0=\sum_{n=1}^N \frac{2i\sigma \sin (\delta\phi_n/2)g_{1,\sigma}}{\cos (\delta\phi_n/2)-i\sigma g_{3,\sigma}\sin (\delta\phi_n/2)}
+\frac{G}{G_q}\frac{i(\varepsilon_{\sigma}+i\delta) g_{1,\sigma}+\Delta g_{3,\sigma}}{\epsilon_{\rm Th}}\label{kr1}.
\end{align} 
We defined $\varepsilon_{\sigma}=\varepsilon+\sigma\mu_B H$. Note, that due to the normalization condition for quasiclassical Green's functions one has $g_{1,\sigma}=\sqrt{1-g^2_{3,\sigma}}$. Due to the small coercivity of EuS the assumption of only one magnetization direction as in \cite{machon:15} is reasonable. This is why the Green's functions decouple in spin space. 

We further assume that all spin mixing angles are the same, $\delta\phi_n=\delta\phi$ and, thus, replace the sum over the channel index $n$ in the matrix current conservation with the number of channels, i.e. $\sum_n\rightarrow N$. However, in a phenomenological way we assume that effectively only a certain fraction $r_N \in[0,1]$ of scattering channels contributes to the spin mixing effect. Alternatively, we may say that the spin mixing only occurs with a certain probability $ r_N $. The corresponding equation then reads
\begin{align}
	0= \epsilon_{\rm Th} r_N N G_Q \frac{2i\sigma \sin (\delta\phi/2)g_{1,\sigma}}{\cos (\delta\phi/2)-i\sigma g_{3,\sigma}\sin (\delta\phi/2)}
	+G\left[i(\varepsilon_{\sigma}+i\delta) g_{1,\sigma}+\Delta g_{3,\sigma}\right]\label{kr2}.
\end{align}
Hence, the strength of the magnetic proximity effect is taken into account in the dimensionless parameter 
$\varepsilon'= r_N (NG_Q/G)(\varepsilon_{\rm Th}/k_B T_c)
=r_N \hbar D N G_Q/\sigma_{el} V k_BT_c = r_N Dk_F^2e^2/8 \pi^2 d \sigma_{el}k_BT_c$.
The physical meaning of the parameter $\varepsilon'$ is understood by using the Fermi gas results for the conductivity $\sigma=e^2N_{\varepsilon_F}D$, the density of states at the Fermi level $N_{\varepsilon_F}=\frac{mk_F}{\hbar^2\pi^2}$, and the Fermi momentum $\hbar k_F=mv_F$. With the definition $\xi_0=\frac{\hbar v_F}{\pi\Delta(T=0)}$ of the superconducting coherence length and the approximation $\Delta(T=0)\approx 1.76\, T_c$, one finds $\varepsilon'\approx 2.76 r_N \xi_0/d$. Hence, the parameter $ \varepsilon' $ determining the influence of the ferromagnetic insulator on the superconducting film becomes smaller for increasing thickness  of the film and with the fraction of spin active channels which is also physically intuitive. 

\begin{figure*}
\centering
(a)\includegraphics[width=0.32\linewidth]{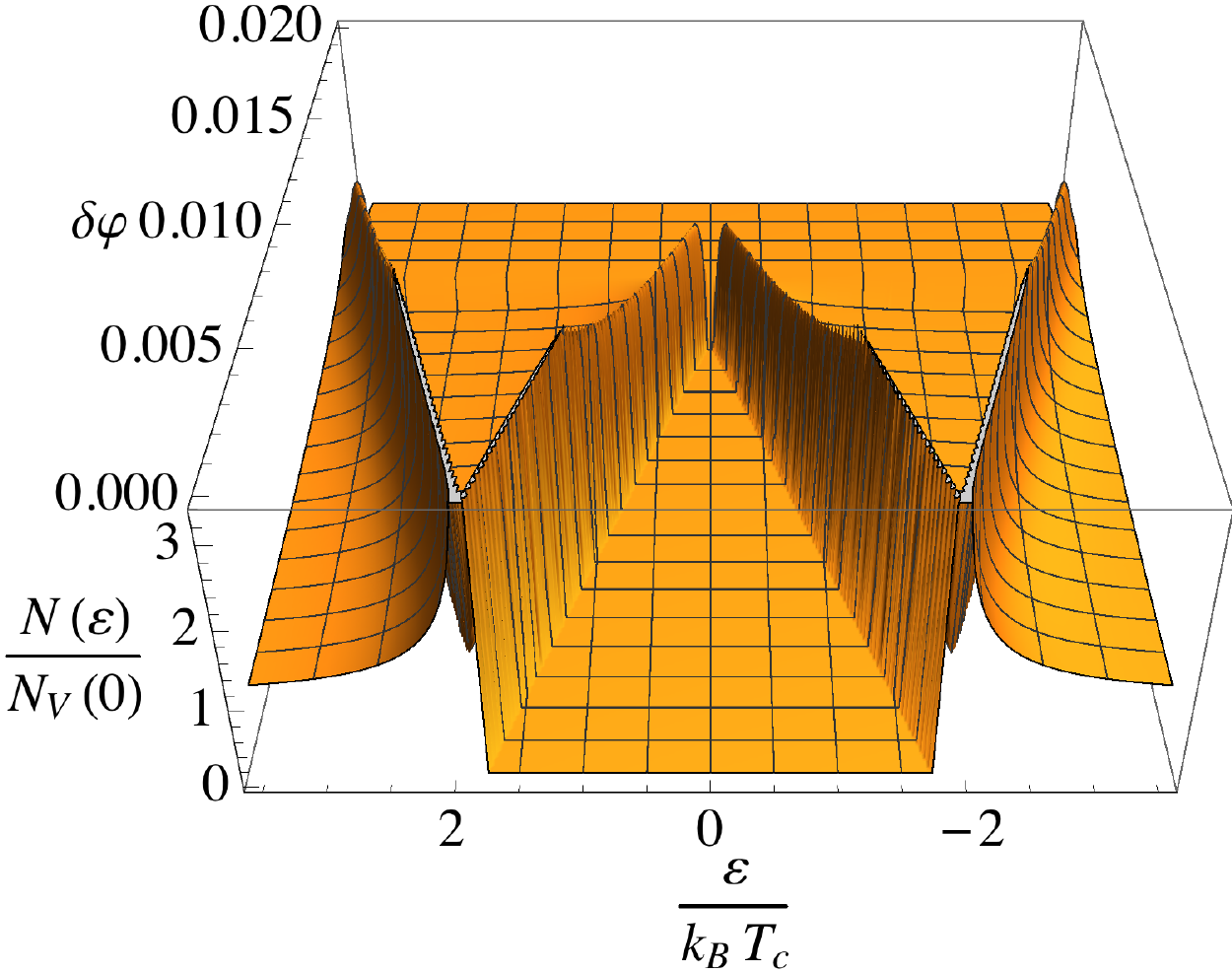}
(b)\includegraphics[width=0.32\linewidth]{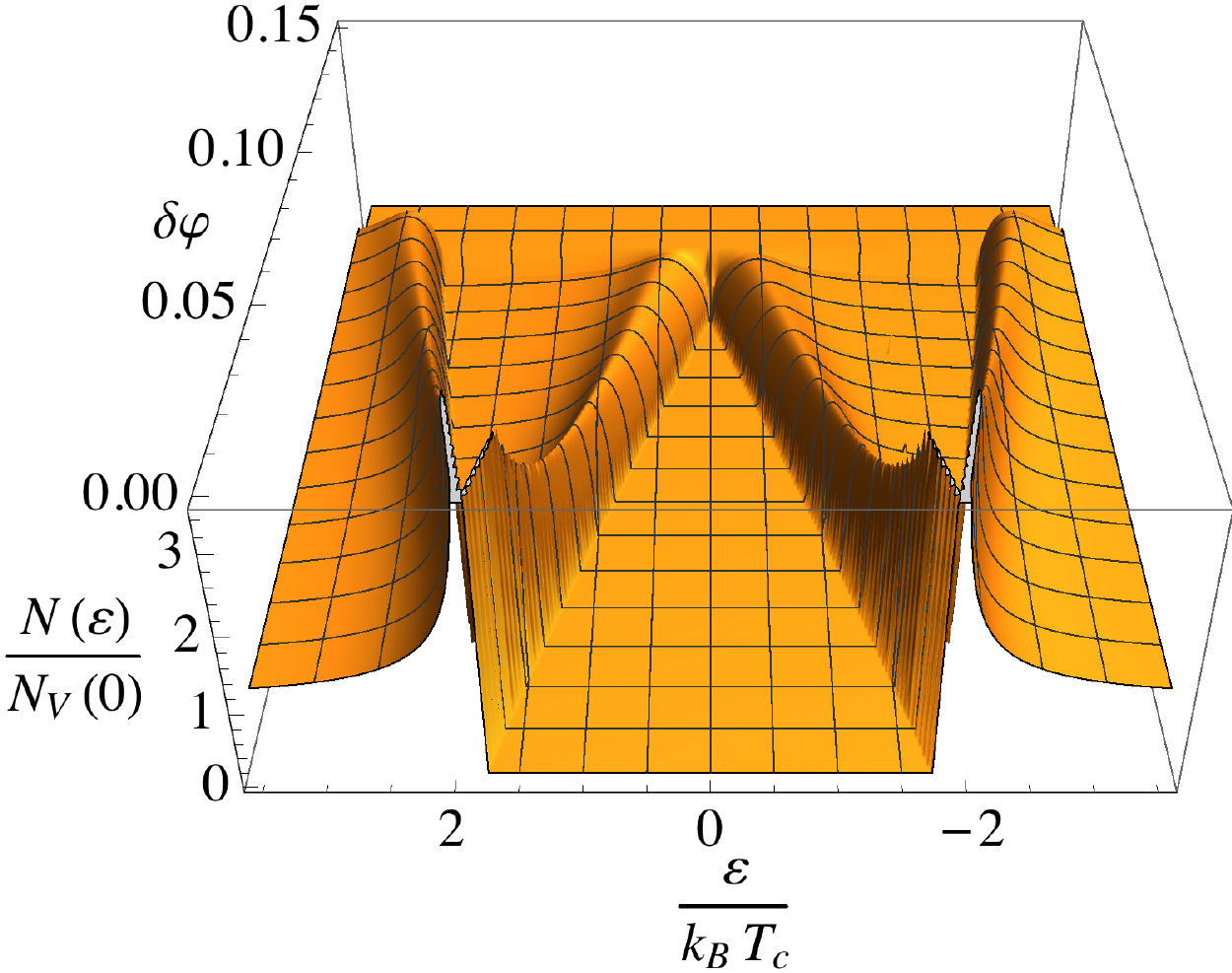}\\
(c)\includegraphics[width=0.32\linewidth]{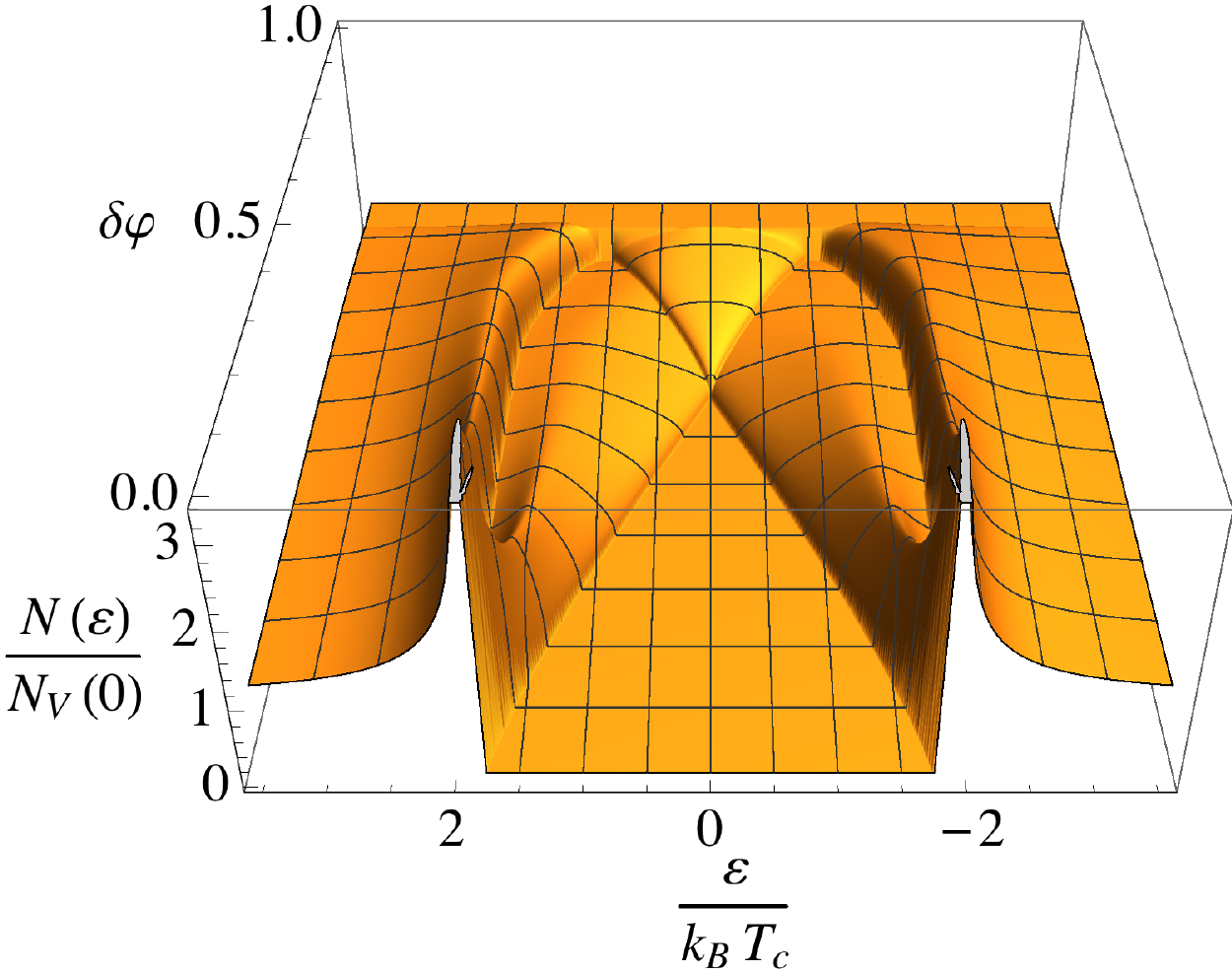}
(d)\includegraphics[width=0.32\linewidth]{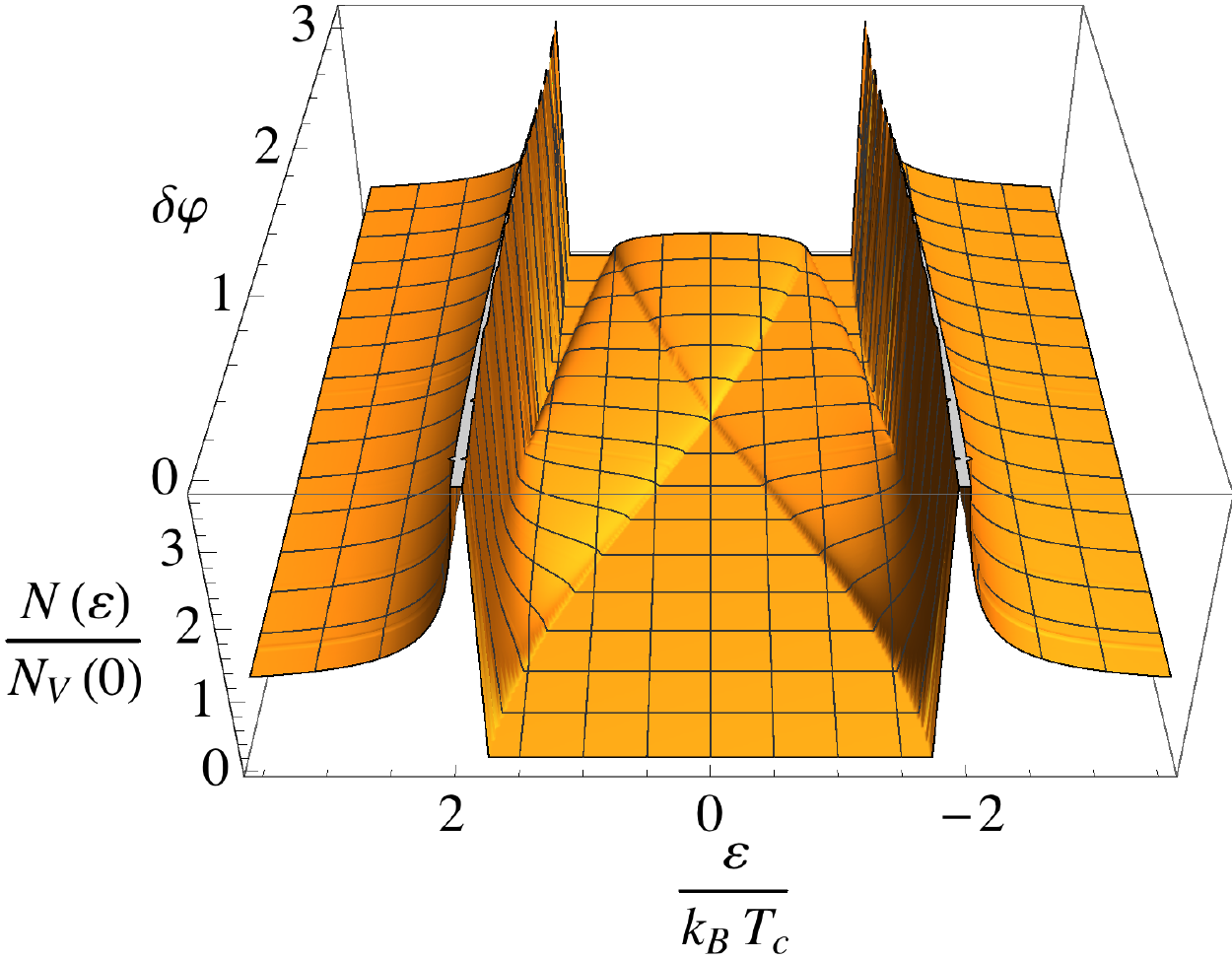}
\caption{\label{dos_sc_fis}
Density of states in a superconductor in proximity to a ferromagnetic insulator indicated by the spin mixing angle $\delta\phi$, with (from (a) to (d)) $\varepsilon'=\{100,10,1,0.1\}$, while the superconducting order parameter is evaluated self-consistently ($T\ll T_c$). 
} 
\end{figure*}

With the above definitions, the BCS self-consistency relation is given by:
\begin{align}\label{fis_bcs} 
\Delta
= \frac{\lambda}{4\pi i}\int_0^{\Omega_{\rm BCS}} d\varepsilon \,\tanh\left(\frac{\varepsilon}{2T}\right)
\Im[g_{1,+}+g_{1,-}] .
\end{align}
We defined the cutoff energy $\Omega_{\rm BCS}$ related to the upper limit of the phonon spectrum. In the following, we use $\Omega_{\rm BCS}=100k_BT_c$., and the coupling constant $\lambda$ that can be eliminated for the bulk superconductor in favor for the critical temperature $T_c$. After solving the fully selfconsistent problem in order to obtain $ \Delta $, the differential tunnel conductance (measured as shown in fig.~\ref{sys_fis}~(a)) is found from the standard definition 
$\frac{dI}{dV}=-G_N\int\nolimits N(\varepsilon)\frac{df}{deV}d\varepsilon$, with the Fermi distribution $f$ and the normal state conductance $G_N$ of the tunnel probe. The density of states is given by $N(\varepsilon)=\frac14\sum_{\sigma\in\{-1,1\}}\Re\rm{Tr}[\hat\tau_3\hat G_{\sigma}]=\frac12\Re(g_{3,+}+g_{3,-})$. Note, that the actual, total density of states per volume is given as $N(\varepsilon)N_{\varepsilon_F}$. 

\section{Phase diagram and density of states}
We now discuss the self-consistency relation for different values of the parameter $\varepsilon'$. Fig.~\ref{gap_fis} shows the phase-diagram for values $\varepsilon'=[100-0.1]$, which for a d=$10$~nm aluminum layer roughly translates into fractions $r_N=[1-0.001]$. The plotted curves are the phase boundaries between superconducting and normal state, while superconductivity is found for temperatures or spin mixing angles beyond these boundaries. The regions with split phase boundaries mark the turnover from second to first order phase transitions, that manifests by the appearance of an additional peak in the free energy and thus of two minima, one at a finite value of $\Delta$ and one at $\Delta=0$. In the solution of the self-consistency problem this thus leads to two finite solutions for $\Delta$ and one at zero. The second solution at finite values is the peak position in the free energy. The second line in fig.~\ref{gap_fis} is the zero crossing of this second finite solution, marking the point when the maximum in the free energy appears or disappears. Hence, the region between the two lines is the instability region where both the superconducting and the normal phase have stable solutions. Besides the different ranges of $\delta\varphi$ in which superconductivity persists, it is seen that also the typical behavior at larger spin mixing angles changes. For small values of $\varepsilon'$, the order parameter depends on $\delta\varphi$ almost exactly as it is expected also to depend on an external exchange field (see e.g. \cite{sarma:63,maki:69}). For thicker films or smaller fractions $r_N$ the instability region becomes smaller. For very small values of $\varepsilon'$ the FI layer can't suppress superconductivity any more and only phase transitions of first order appear. Note, that everything calculated here is $2\pi$-periodic in $\delta\varphi$ and hence at $\delta\varphi=\pi$ the maximum effect is already reached.

Now, we discuss the density of states dependence on the spin mixing angle $\delta\phi$ for different values of the parameter $\varepsilon'$. For the sake of clearness the Zeeman splitting from the external field is ignored at this point. The external field is rather taken into account to create stronger magnetization in the FI layer that is assumed to be proportional to the spin mixing angle. Later, this simplification will be justified by showing that the effect of the Zeeman splitting from the external field is orders of magnitude smaller than the effects from the internal magnetization.

The changes in the density of states of the superconductor are dominated by two effects. On one hand the initial peaks at the $T=0$ superconductor gap $\Delta_0$ are spin-split into two separated peaks, each positioned depending on $\delta\varphi$ and $\Delta$. On the other hand $\Delta$ self-consistently also depends on the spin mixing angle.

In fig.~\ref{dos_sc_fis}, we plot the density of states for $T\ll T_c$ with self-consistent $\Delta$. For very thin layers ($\varepsilon'=100$) the peaks (initially at $\Delta$) symmetrically split into their spin components. This behavior is also observed in external fields, as already measured e.g. in \cite{tedrow:71}. However, with increasing layer size or fraction $r_N$, the superconductivity persists for larger spin mixing angles and the behavior gets nonlinear until a completely different situation is found at $\varepsilon'=0.1$. Here, the outer peak position nearly stays while only the inner peak moves towards (and across) the Fermi level. Another effect of larger spin mixing angles is that the inner peak is broadened, and finally becomes a wide and flat band. Besides this, the self-consistency relation for thin films produces the typical step-like first order phase transition at the critical $\delta\phi$ (here always plotted for the upper branch of fig.~\ref{gap_fis}), while especially in the case $\varepsilon'=1$ a significant shift of the peak positions is visible for larger spin mixing angles. For thick enough layers (or small portions $r_N$) the effect of the spin mixing on $\Delta$ almost becomes not relevant.

\section{Comparison of experiment and theory}

\begin{figure}
\centering
\includegraphics[width=0.8\linewidth]{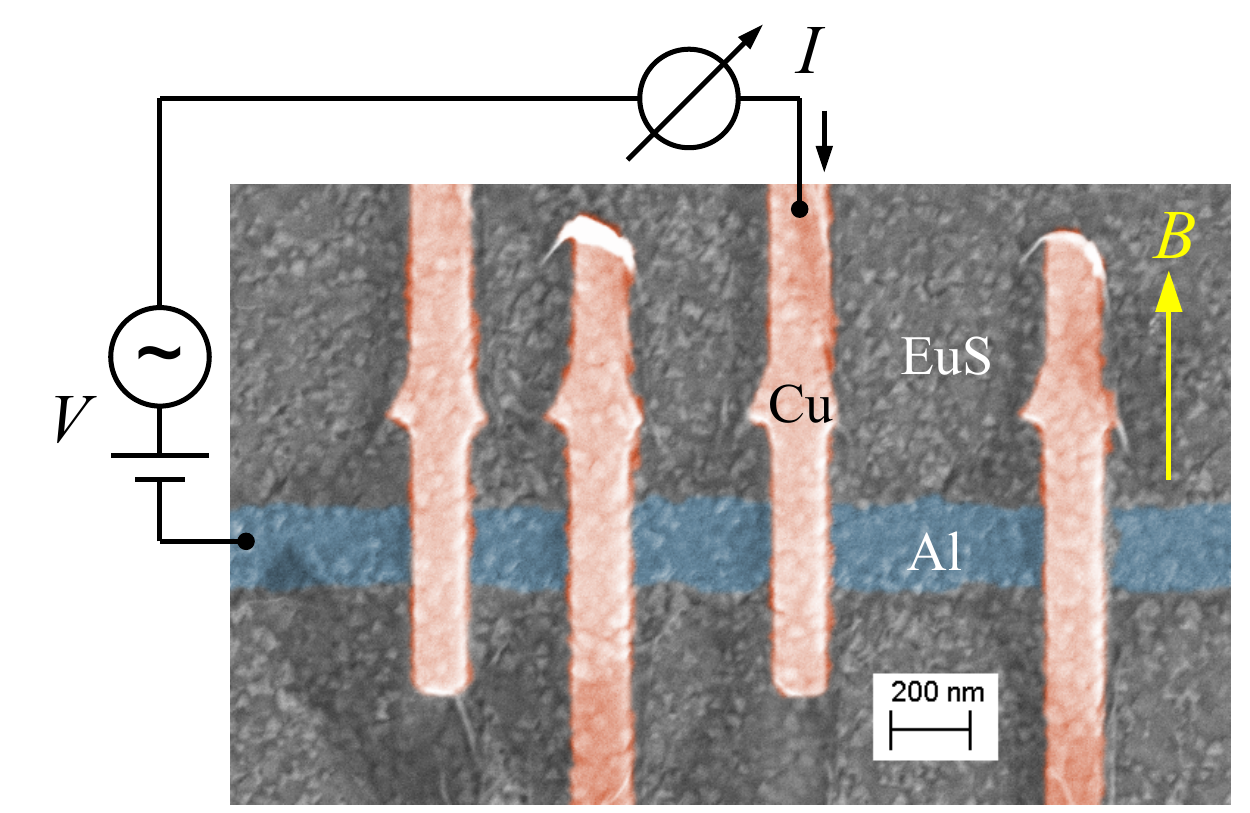}
\caption{\label{fig_sample}
False-color scanning electron microscopy image of the sample, and experimental scheme.} 
\end{figure}

To illustrate our model, we use it to fit experimental data obtained on a sample made of a superconducting aluminum film on top of the ferromagnetic insulator europium sulfide. Fig.~\ref{fig_sample} shows a false-color scanning electron microscopy image of the sample, together with the experimental scheme. The sample was fabricated in a two-step procedure: first, a EuS film of $44~\mathrm{nm}$ thickness was created by e-beam evaporation of EuS onto a Si 111 substrate heated to $800~^\circ\mathrm{C}$. In a second fabrication step, aluminum/aluminum oxide/copper tunnel junctions were fabricated on the EuS film using e-beam lithography and shadow evaporation. The nominal aluminum film thickness was $d=10~\mathrm{nm}$. The differential conductance $g=dI/dV$ of the tunnel junctions was measured as a function of bias voltage $V$ using standard low-frequency lock-in techniques in a dilution refrigerator as base temperatures down to $T=50~\mathrm{mK}$ with an in-plane magnetic field $B$ applied along the direction of the copper wires, as indicated in Fig.~\ref{fig_sample}. Details of the film fabrication, magnetic properties, and experimental procedures can be found in \cite{wolf2014b,wolf2014c}.

\begin{figure}
\centering
\includegraphics[width=\linewidth]{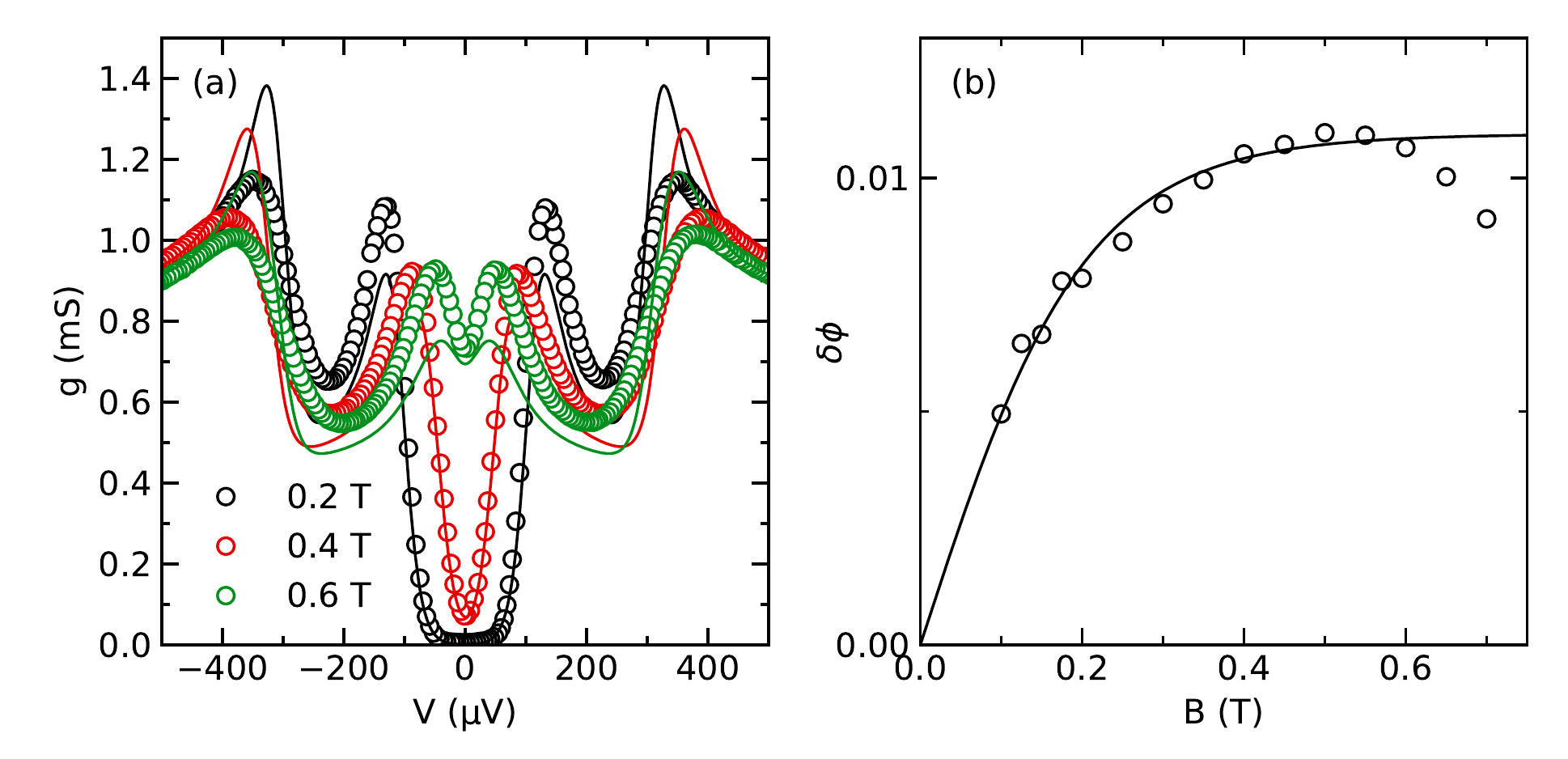}
\caption{\label{fig_results}
(a) differential conductance $g$ as a function of bias $V$ for different applied magnetic fields $B$ (symbols), and fits with our model (lines). (b) Spin-mixing angle $\delta\phi$ as a function of the applied field $B$ extracted from the fits (symbols), and fit with a Brillouin function (line).} 
\end{figure}

Examples of the conductance spectra measured for different applied fields in one of the junctions is shown in Fig.~\ref{fig_results}(a). At small fields, the spectra exhibit a well-defined gap with negligible subgap conductance, indicating a defect-free tunnel barrier. A spin splitting of the density of states is clearly visible. The observed splitting greatly exceeds the expected splitting due to the Zeeman energy $\epsilon_\mathrm{Z}=\mu_\mathrm{B}B$ (which is about $35~\mathrm{\mu eV}$ at $B=0.6~\mathrm{T}$). The solid lines in Fig.~\ref{fig_results}(a) are fits with our model. We have included orbital depairing in the fits, with an orbital depairing parameter \cite{maki1964a}
\begin{equation}
 \alpha_\mathrm{orb} = \frac{1}{2}\left(\frac{B}{B_\mathrm{c,orb}}\right)^2
\end{equation}
for a thin film in an in-plane field. From known sample parameters we estimate $B_\mathrm{c,orb}\approx 2~\mathrm{T}$ and $\epsilon^\prime \approx 70$, which leaves us with $\Delta$ and $\delta\phi$ as free parameters. The fits give a good account of the observed spin splitting. The spin-mixing angle extracted from the fits is plotted in Fig.~\ref{fig_results}(b). It is found to depend on the applied magnetic field over the entire field range. In contrast, the EuS magnetization is saturated above a few mT in our film \cite{wolf2014b}. A similar dependence of the spin splitting on the applied field is a commonly observed in EuS/Al structures \cite{hao1990,xiong2011}, and the microscopic origin is yet unclear. A possible explanation are misaligned spins at the interface, which are nearly free and therefore gradually aligned by the applied field. The misaligned spins might be the result of partial oxidation of the EuS surface during sample transfer between our two fabrication steps. Lacking a microscopic model, we have attempted to fit the field dependence of $\delta \phi$ with a Brillouin function. The fit is shown as a line in Fig.~\ref{fig_results}(b). It is in reasonable agreement with the data up to about 0.6 T, with an effective angular momentum $J\approx 0.74\hbar$. Above 0.6 T, the data deviate downwards from the fit, and these data points were excluded from the fit. The deviation can be explained by Fermi-liquid renormalization of the effective spin splitting near the critical field \cite{alexander1985,xiong2011}, which is not included in our model.

\section{Conclusions}
Concluding, based on the boundary condition \cite{machon:15} augmenting the circuit theory \cite{nazarov:99}, we investigated FI-S heterostructures in the dirty limit. 
We discussed the dependence of the density of states (and thus also the differential conductance) on the spin mixing angles for different layer thicknesses observing strong deviations from the typically linear behavior in exchange fields. The model displays a new phase-diagram that strongly depends on the spin mixing angle and includes a crossover from a first- to a second-order phase transition.

We applied our theory to our experiment measuring the differential conductance in an EuS-Al bilayer. In the experiment, an enhanced spin splitting of the density of states in an external magnetic field was observed. To reproduce the experimental data, we have determined the spin-mixing angle as a function of the applied magnetic field, and given an estimate is given on how to take into account the relation between the external field and the spin mixing angle. 

We are thus confident that our theory will in the future provide further  motivation for the interesting physics of ferromagnetic insulators and the proximity effect in ferromagnet or antiferromagnet - superconductor heterostructures.

\bibliography{EuS_DoS_V6}

\end{document}